# Experimental and Theoretical Exploration of Terahertz Channel Performance through Glass Doors

Da Li[1], Wenbo Liu[1], Menghan Wei[1], Jiacheng Liu[1], Guohao Liu[1,2], Peian Li[1], Houjun Sun[1,2,3], Jianjun Ma[*,1,2,3]

[1]School of Integrated Circuits and Electronics, Beijing Institute of Technology, Beijing, 100081 China
[2]Tangshan Research Institute Beijing Institute of Technology Tangshan, Hebei, 063099 China
[3]Beijing Key Laboratory of Millimeter and Terahertz Wave Technology, Beijing, 100081 China

*jianjun_ma@bit.edu.cn

**Abstract**

In the evolving landscape of terahertz communication, the behavior of channels within indoor environments, particularly through glass doors, has garnered significant attention. This paper comprehensively investigates terahertz channel performance under such conditions, employing a measurement setup operational between 113 and 170 GHz. Analyzing scenarios frequently induced by human activity and environmental factors, like door movements, we established a comprehensive theoretical model. This model seamlessly integrates transmission, reflection, absorption, and diffraction mechanisms, leveraging the Fresnel formula, multi-layer transmission paradigm, and knife-edge diffraction theory. Our experimental results and theoretical predictions harmoniously align, revealing intricate dependencies, such as increased power loss at higher frequencies and larger incident angles. Furthermore, door interactions, whether opening or oscillations, significantly impact the terahertz channel. Notably, door edges lead to a power blockage surpassing the transmission loss of the glass itself but remaining inferior to metallic handle interferences. This paper's insights are pivotal for the design and fabrication of terahertz communication systems within indoor settings, pushing the boundaries of efficient and reliable communication.

## Introduction

The advent of terahertz (THz) technology heralds a paradigm shift in wireless communication, addressing the escalating demand for higher data transmission rates spurred by the burgeoning proliferation of data as a new axis in production [1, 2]. A forecast posits a convergence of wireless and wired communication rates, an evolution where THz technology is anticipated to play a pivotal role [3, 4]. With its expansive bandwidth, the THz spectrum is poised to unlock unprecedented data transmission speeds, potentially reaching the Tbps echelon [5, 6].

However, the propagation characteristics of THz waves present unique challenges. Compared to conventional wireless frequencies, THz waves exhibit significantly higher attenuation and reduced transmission distances [7-9]. Consequently, indoor environments emerge as prime candidates for THz communication applications [10]. Within these confines, non-line of sight (NLOS) paths, predominantly formed through reflection, become integral components of the received signal power [11]. An in-depth exploration of THz reflection properties is crucial, offering insights to enhance the signal-to-noise ratio at the physical layer. Transmission, another predominant mode of THz propagation, commands attention for its potential to minimize propagation losses and secure THz communications [12]. Scattering within the THz channel, particularly in glass mediums, introduces signal attenuation and potential security vulnerabilities [13]. This phenomenon is prevalent in various scenarios, such as indoor-to-outdoor communications, vehicle-to-vehicle interactions and vehicle-to-infrastructure data connections *etc*. The present study delves into the impact of glass edges and door handles on signal behavior, shedding light on the propagation characteristics of THz channels through glass mediums, and underscoring the importance of this investigation for indoor THz channel modeling.

The diversity in propagation characteristics across the extensive THz frequency range necessitates a foundational understanding of these traits [14]. Beyond the ambit of 5G networks, THz frequencies are envisioned to revolutionize short-range communications, potentially through ultra-dense deployments of access points within a 10-meter radius [15]. The focus of indoor THz channel modeling thus gravitates towards commonplace indoor settings such as offices. Prior studies have contributed to this knowledge pool, examining path loss, time delay, and angular spread across various indoor scenarios [15-17]. These investigations utilize sophisticated experimental setups, encompassing RF fronts, horn antennas, and vector network analyzers (VNA), complemented by rotatable platforms to manipulate elevation and azimuth angles. Through these experiments, intricate relationships between signal power, time delay, and angular spread have been established. The presented path-loss heatmaps, derived from four distinct propagation models (ITU, LD, FS, and COST 231), offer comparative insights and guidelines for model application across varied scenarios [18]. Furthermore, large-scale path loss results from indoor office settings reveal remarkable consistency across multiple frequencies, highlighting similarities between THz channels and existing millimeter-wave propagation channels [19-22]. In the realm of nano-networks within the THz band, models accounting for small particle scattering effects have been developed, enriched by comprehensive measurements across different antenna types and transceiver locations [23]. The THz band's behavior in terms of diffraction phenomena has also been scrutinized, with extensive studies conducted on various shapes and materials [24, 25]. For short-range indoor communication, the spatial and temporal characteristics of the THz band have been analyzed, encompassing specific scenarios such as device-to-device desktop communication channels in multipath fading conditions [26-28].

Despite these advancements, the comprehensive impact of glass doors on THz channels has remained largely unexplored until now. Previous research has provided valuable insights, with one study measuring the penetration loss of a 300 GHz signal through float glass [29], and another extending the exploration to both reflection and transmission characteristics over a 220-320 GHz frequency range using a Vector Network Analyzer (VNA) and a sophisticated simulation model [30]. This paper however, ventures into previously uncharted territory by drawing upon experimental data to dissect the reflection and transmission characteristics of THz waves through glass in indoor settings, over the 110-170 GHz frequency range. We scrutinize the effects of glass door movements on THz communication channels, employing the double knife-edge diffraction model to characterize signal attenuation. Furthermore, we explore the alterations in communication channels attributed to glass door vibrations after door opening, unveiling the correlation between vibration amplitude and channel degradation. The culmination of this research synthesizes experimental findings and theoretical models, offering a comprehensive understanding of THz channels through glass doors in indoor scenarios.

## II. Experimental setup

To delve into the nuances of signal strength, we established a non-modulated THz channel as depicted in Fig. 1. The transmission apparatus comprises a signal generator (Ceyear 1465D), a frequency multiplier module (Ceyear 82406B), and a horn antenna (HD-1400SGAH25) augmented by a dielectric lens with a focal length of 10 cm. The Ceyear 1465D is capable of generating signals up to 20 GHz, subsequently up-converted to the 113-170 GHz range by the Ceyear 82406B. Continuous wave (CW) signals, spanning frequencies from 113 to 170 GHz, are radiated from the horn antenna for the measurements. On the receiving side, an identical horn antenna captures the signal, routing it to a power sensor (Ceyear 71718) for detection. A laptop controls the signal generator's frequency, with a frequency step of 1 GHz. The alignment between the transmitting and receiving antennas is precisely centered across the width of the glass door.

The glass door (265.5 cm×113.5 cm×1.2 cm in L×W×T) under investigation is constructed of standard glass and situated in the gymnasium of the Liangxiang Campus at Beijing Institute of Technology (BIT). The door features metal handles placed 10.3 cm from its edge, each spanning 1.72 m in length and 4.1 cm in width. The ensemble includes four doors aligned in a row, each capable of opening at various angles. The glass utilized in our experiments shares similarities with that described in reference [31], affirming our belief that the models and methods derived from this work are transferable to analogous scenarios in the other countries.

To examine the reflectivity at the surface of the glass door with precision, a slender layer of aluminum was applied to overlay the glass, a method that promotes specular reflection, as referenced in studies [32, 33]. The immaculate surface of the aluminum, a result of its attachment to the glass, ensures an extraordinary level of flatness, with surface roughness being incredibly small (should be < 20 μm compared with our previous measurements in reference [32]). This minimal surface roughness ensures that losses attributed to this factor remain inconsequential for frequencies below 1 THz. Given these attributes, the aluminum plate stands out as an exemplary reflector within the terahertz frequency spectrum, making it highly suitable for our reflective experimentation [32].

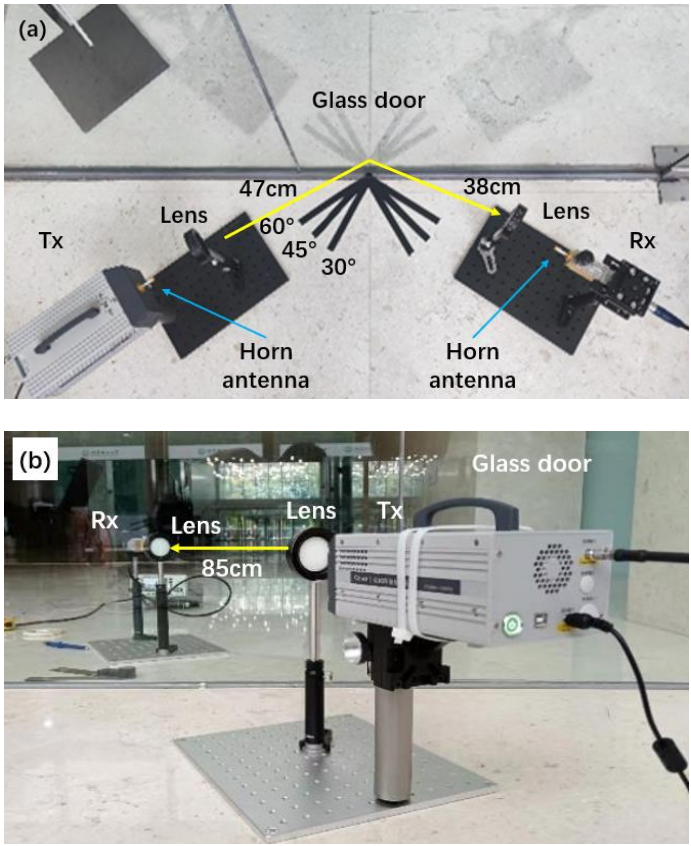

Figure 1 Pictures of the setups for (a) reflectivity measurement and (b) transmission measurement.

### III. Measurement and Calculation Results

### 3.1 Interaction with a single glass door

This sub-section elucidates the phenomena of terahertz wave transmission and reflection as it interacts with a monolithic glass layer. The experimental setup was calibrated to observe interactions within the frequency range of 113 to 170 GHz, and incidence angles set at 0°, 30°, and 60° for transmission; and 30°, 45°, and 60° for reflection. The frequency sweep conducted in both experimental conditions was incremental, with a step size of 1 GHz. As demonstrated in Fig. 2(a) and (b), a direct relationship between the incident angle and reflectivity is observed, alongside an inverse relationship with transmittance. As frequency increases, reflectivity undergoes minimal alteration, whereas transmittance demonstrates a consistent decrease. It is postulated that this decrease in transmittance is attributed to the uniform absorption rate of electromagnetic waves by the glass across the entire frequency range under investigation, which has been confirmed in reference [34] by measured data.

In our theoretical model, as depicted in Fig. 2(c), the terahertz wave, upon entering the glass medium, undergoes multiple internal reflections. This model, grounded in the Fresnel formula, enables the computation of reflection and transmission coefficients across various frequencies and incidence angles, culminating in the determination of reflectivity and transmittance. The model employs the Fresnel formula for parallel polarized waves (used in our setup), integrating parameters such as wave impedances in the two mediums, and the dielectric constants of air and the glass layer. The dielectric constant utilized for computations is derived from reference [31].

In the single-layer terahertz glass transmission & reflection model (Fig. 2(c)), the system comprises three layers: air, -glass, and -air. Given the occurrence of multiple internal reflections and transmissions within the sample, the total reflection coefficient ($\rho_s$) and transmission coefficient ($\tau_s$) are expressed as follows:

$$\rho_s = \rho_{12} + \frac{\tau_{12}\rho_{23}\tau_{21}L^2}{1 - \rho_{21}\rho_{23}L^2} \tag{1}$$

$$\tau_s = \frac{\tau_{12}\tau_{23}L}{1 - \rho_{21}\rho_{23}L^2} \tag{2}$$

with the transmission loss in glass layer $L = e^{-\gamma d \cos\theta_2}$. The propagation constant ($\gamma$) can be $\gamma = \alpha + i\beta$, where $\alpha$ denotes the attenuation constant in the glass layer, $\beta$ represents the phase constant. According to Fresnel's formula for parallel polarized waves, the reflection coefficient $\rho_{ij} = \frac{\eta_j \cos\theta_i - \eta_i \cos\theta_j}{\eta_j \cos\theta_i + \eta_i \cos\theta_j}$ and transmission coefficient $\tau_{ij} = \frac{2\eta_j \cos\theta_i}{\eta_j \cos\theta_i + \eta_i \cos\theta_j}$ from the *i*-th layer medium to the *j*-th layer medium can be obtained with $\theta_i$ as the incident angle and $\theta_j$ as the angle of refraction. Parameters $\eta_k = \frac{\eta_0}{\sqrt{\varepsilon}}$ (when $k$ is 1 or 3, $\varepsilon$ is $\varepsilon_1$; when $k$ is 2, $\varepsilon$ is $\varepsilon_2$) is the wave impedance in medium respectively, and $\eta_0$ is the wave impedance in free space, parameters $\varepsilon_1$ and $\varepsilon_2$ are the dielectric constants of air and glass layers, respectively.

A comparative analysis of calculated curves and experimental data reveals a strong correlation, particularly evident in Fig. 2(a), showcasing theoretical transmission attenuation. The attenuation model illustrates an increase proportional to frequency, aligning closely with the measured data. It is noteworthy that theoretical attenuation is contingent upon the incident angle, with larger angles resulting in increased attenuation. This trend finds congruence with findings in reference [4].

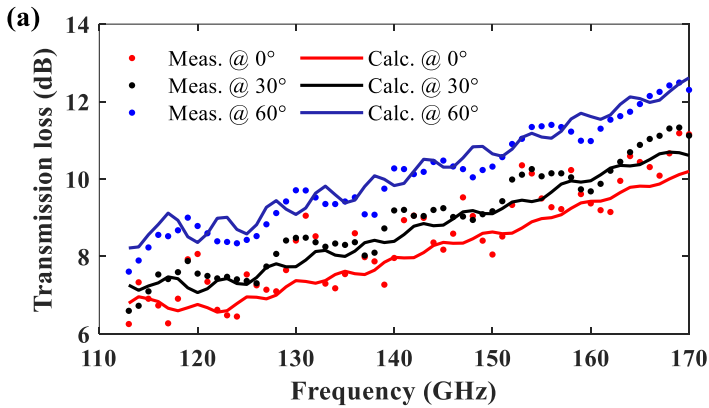

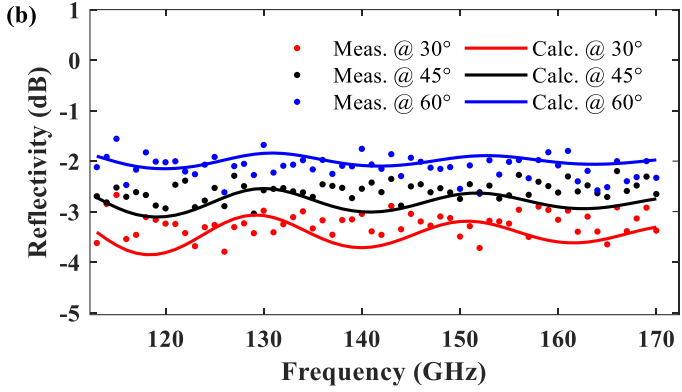

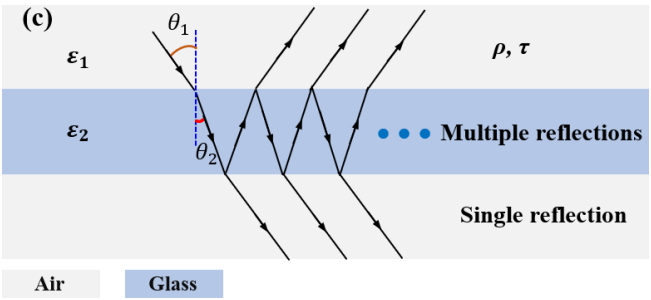

Figure 2(a) Transmission attenuation at 0, 30, 60 degrees of incidence at 113-170 GHz; (b) Reflectivity at 30, 45, 60 degrees of incidence at 113-170 GHz; (c) Single-layer medium model.

## 3.2 Transmission through dual glass doors

Considering the architectural preference for multi-layered glass structures to enhance thermal insulation and energy efficiency [35], we extends our focus to the transmission of terahertz waves through dual glass doors (which cannot be extend to more glass layers due to the power limitation of our setup). As illustrated in Fig. 3(a), the experimental setup was meticulously arranged, maintaining a 5.62 m distance between the transmitter and receiver, and situating the antennas at a height of 1.02 m. The experimental array included four doors (D1-D4), each situated at specific distances from the receiver. A singular door was engaged at a time, and each door was systematically opened to assess glass transmission at varying frequencies.

The transmission power through each door was meticulously recorded, averaged, and presented with error bars in Fig. 3(b). The consistency in error bar length reaffirms that the door's position within the channel path does not significantly impact transmission loss. Theoretical predictions employing Eq. (2) align remarkably well with the measured data, despite the presence of oscillations not captured in the measurements – a discrepancy attributed to inhomogeneities [36] in the glass layers.

When assessing the transmission through two glass doors, configurations such as D1-D2, D1-D3, D1-D4 and D2-D3 were examined, with the averaged values and error bars once again depicted in Fig. 3(b). Beyond the 145 GHz threshold, detected channel power plummeted below noise levels, indicative of substantial power losses. Notably, the attenuation due to two layers of glass doors proved less than double the loss incurred by a single layer, challenging the validity of simplistic additive algorithms for multi-layer prediction.

The dual-layer THz glass transmission model, depicted in Fig. 3(c), comprises five layers: air, -glass, -air, -glass, and -air. Given the substantial distance between the doors, reflections by interface between the 3-4 layers, as well as that by between the 4-5 layers, are deemed negligible. The total transmission coefficient ($\tau_m$) is computed as follows:

$$\begin{aligned}
\tau_m &= \tau_{12}L_2L_3L_4\tau_{23}\tau_{34}\tau_{45} + \tau_{12}L_2^3L_3L_4\rho_{23}\rho_{21}\tau_{23}\tau_{34}\tau_{45} + \tau_{12}L_2^5L_3L_4\rho_{23}^2\rho_{21}^2\tau_{23}\tau_{34}\tau_{45} + \cdots \\
&= \tau_{12}L_2L_3L_4\tau_{23}\tau_{34}\tau_{45}(1 + \rho_{21}\rho_{23}L_2^2 + \rho_{23}^2\rho_{21}^2L_2^4 + \cdots) \\
&= \frac{\tau_{12}L_2L_3L_4\tau_{23}\tau_{34}\tau_{45}}{1-\rho_{21}\rho_{23}L_2^2}
\end{aligned} \quad (3)$$

where transmission losses in layers two, three, and four are symbolized by $L_2 = e^{-\gamma d_2 \cos\theta_2}$, $L_3 = e^{-\gamma d_3 \cos\theta_1}$, $L_4 = e^{-\gamma d_4 \cos\theta_2}$, respectively. The predicted curve, while displaying oscillations due to multi-reflection induced interference, aligns closely with measured data below 145 GHz, underscoring the efficacy of Eq. (3) for characterizing channel behavior through multi-layer glass structures. By comparing the black curve (double of the calculated power loss caused by a single layer) and the read curve (calculated power loss caused two layers of glass doors), there is around a 3-dB discrepancy, which further confirms that the power attenuation resulting from the interaction with dual layers of glass doors was discovered to be less than twice the loss experienced when engaging with a singular layer. The efficacy of straightforward, cumulative methodologies for the anticipation of multi-layer effects should be checked in future further channel modeling.

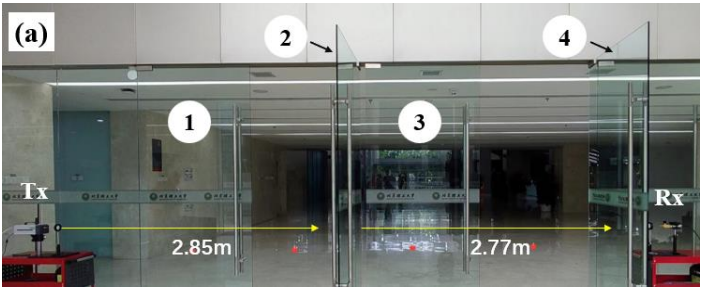

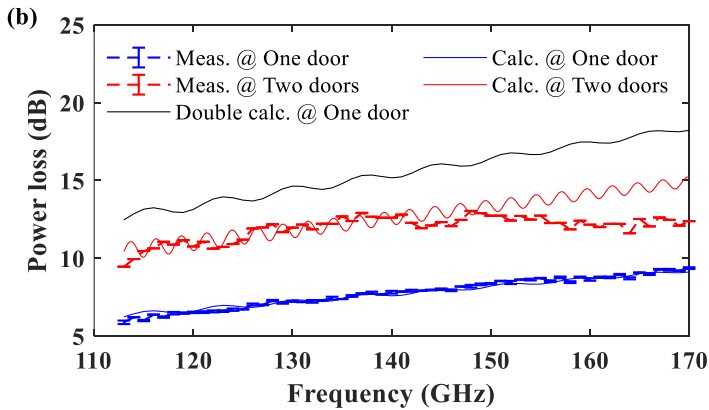

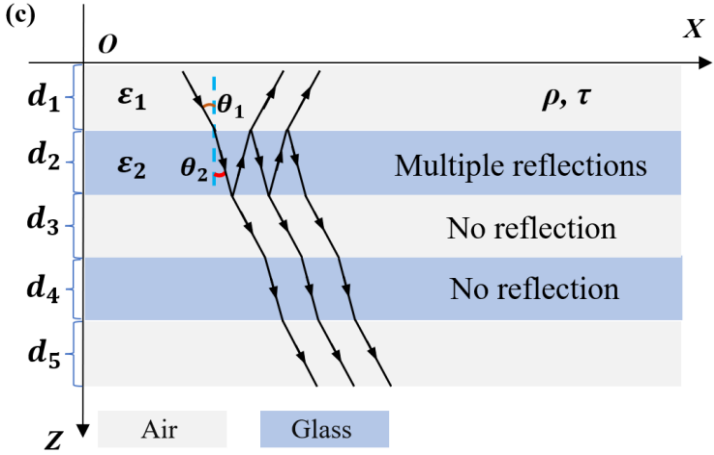

Figure 3 (a) Schematic diagram of the transmittance measurement through several glass doors; (b) Measured and calculated channel power degraded through one and two layers of glass doors; (c) Two-layer medium model [37].

### 3.3 Channel degradation due to opening glass door

The transition from indoor to outdoor environments, or for the vehicle-to-vehicle (V2V) communications, requires an intricate understanding of the terahertz channel's behavior [38, 39]. The act of a glass door opening introduces a scattering and blockage phenomenon that is pivotal for the attenuation analysis of the terahertz channel. This is instrumental in devising strategies to augment the channel's reliability and bolster signal reception. The experimental configuration to measure the attenuation induced by the glass door's edge and handles is depicted in Fig. 4(a). Here, the transmitter and receiver are strategically placed over a distance of 1.7 m, flanking the door on either side. The system operates at a frequency of 140 GHz, selected for its propensity to yield a high output power from the transmitter and potential for future communication systems [9, 40]. Over a span of 23 seconds, measurements were meticulously recorded, ensuring a comprehensive capture of the door's entire opening trajectory.

Upon examination of the data presented in Fig. 4(b), the identification of three distinct peaks in signal loss is apparent. These correspond to the diffraction events incurred by the door's edge and its two handles during the act of opening. The power dissipation attributed to the door's edge stands at a significant 10 dB, surpassing the ~8 dB loss encountered during glass transmission as in Fig. 2(a), yet falling short of the ~15 dB diffraction loss induced by the metallic handles. It is pertinent to note that this analysis does not account for blockage by human body [15], as its impact (more than 40 dB loss) exceeds the power margin of our current measurement apparatus.

To prognosticate these observations, we engaged the double knife-edge diffraction model (DKED), a methodology validated in its efficacy for human blockage scenarios in reference [15]. This approach amalgamates four edges, each delineated through a simplified knife-edge diffraction model as elucidated in reference [41]. The ensuing power loss, resultant from obstructions posed by the door's edge or handles, is computed via the expression

$$L_{\text{diff}}[\text{dB}] = -20 \log_{10}(1 - (F_{w1} + F_{w2})) \qquad (4)$$

with parameters $F_{w1}$ and $F_{w2}$ representing the shadowing losses attributed to both edges of the obstructing entity. The theoretical underpinnings of this calculation are grounded in the expression

$$F_{w1|w2} = \frac{\tan^{-1}(\pm\frac{\pi}{2}\sqrt{\frac{\pi}{\lambda}(D2_{w1|w2}+D1_{w1|w2}-r)})}{\pi} \tag{5}$$

where variables $D1_{w1}$ and $D2_{w1}$ denote the distances from the transmitter and receiver to the w1 edge of the object, respectively, while variables $D1_{w2}$ and $D2_{w2}$ correspond to the w2 edge. The parameter $r$ encapsulates the distance interlinking the transmitter and receiver, and $\lambda$ symbolizes the wavelength. Here, the "+" sign is indicative of the NLOS path, while the "-" sign represents the LOS (line of sight) path. For our experimental setup, the "+" sign is adopted, correlating to the blockage incurred by the door's edge and handles. As illustrated in Fig. 4(b), there exists a commendable concordance between the calculated curves and the experimentally obtained data at instances where blockage is evident. Furthermore, it is imperative to highlight that Eq. (4) demonstrates proficiency in prognosticating the power evolution surrounding the peak of blockage-induced losses, underscoring its potential applicability in time-variant channel characterization studies.

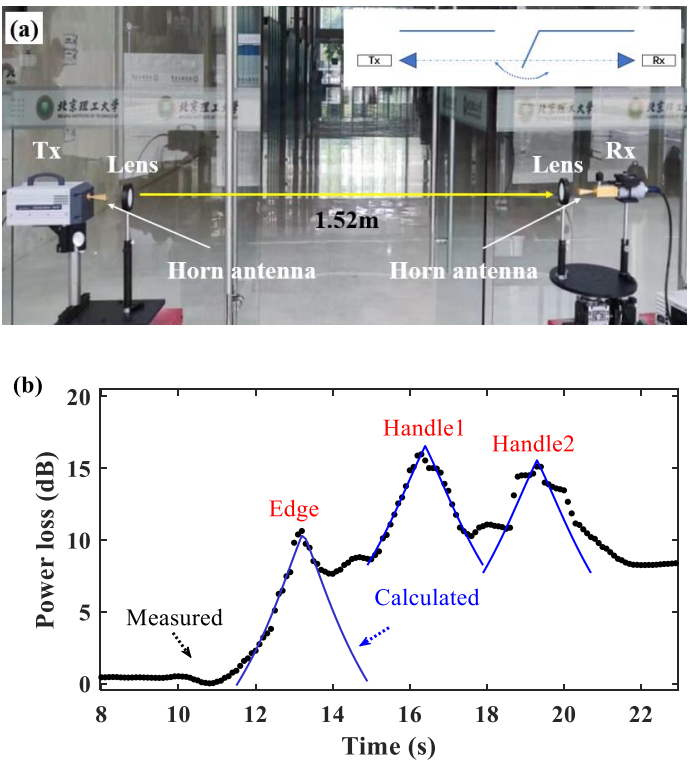

Figure 4(a) Picture of channel measurement setup with door opening. (b) Measured and calculated channel power loss due to door opening.

### 3.4 Channel impairment due to glass door oscillations

The dynamic interactions between indoor and outdoor environments [42], particularly when mediated through glass doors, elicit changes in the terahertz channel, necessitating a comprehensive exploration. In this segment, we delve into deciphering the correlation between the amplitude of door oscillations and the consequential alterations in the terahertz channel. The experimental setup is meticulously configured with the transmitter and receiver placed perpendicular to the surface of the glass door, as illustrated in Fig. 5(a), operating at a frequency of 140 GHz. The glass door is subjected to oscillations of varying amplitudes to simulate real-world door movements, with the received signal's power fluctuations being the focal point of our measurements.

The ensuing data, depicted in Fig. 5(b), showcases the power evolution across different time intervals and oscillation amplitudes of 5 cm, 10 cm, and 15 cm. The amplitude of oscillation, denoted as σ, is quantified as the maximal displacement exhibited by the door's edge, as showcased in the inset of Fig. 5(a). The results manifest a clear periodic trend in power evolution across all three oscillation conditions, with the periodicity being contingent on the amplitude of oscillation. It is discernible that the power variations are directly influenced by the amplitude of door oscillations, as elucidated in the inset of Fig. 4(b). The Root Mean Square (RMS) value of power variations escalates concomitantly with an increase in oscillation amplitude, signifying that door movements, even those smaller than 0.2 dB, can instigate measurable fluctuations in the terahertz channel. However, it is crucial to underscore that these fluctuations do not translate into pronounced channel degradation [28], ensuring the resilience of the terahertz channel under such dynamic conditions in our measurements.

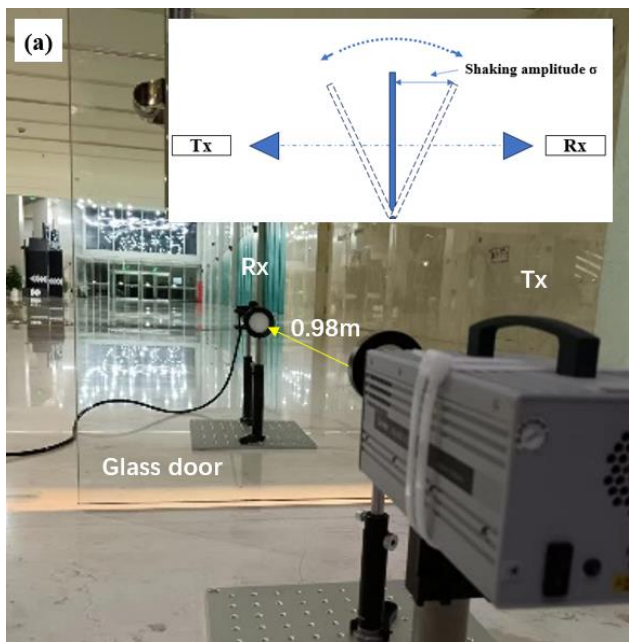

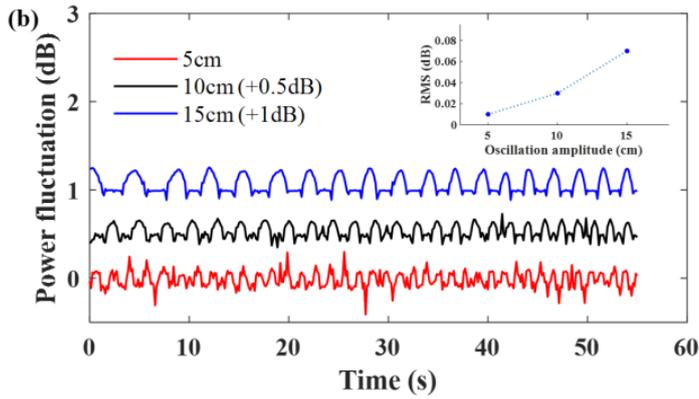

Figure 5(a) Picture of channel measurement setup in the case of glass door oscillation; (b) Temporal variation of the measured channel power due to glass door oscillation.

**Conclusion**

In the realm of terahertz communications, understanding channel performance within indoor environments is of paramount importance, particularly when it involves interactions with glass doors. This knowledge is crucial for the meticulous design of communication prototypes and the precise fabrication of systems. This paper delves into the intricacies of terahertz channels traversing through glass doors, employing a comprehensive measurement setup with operational frequencies ranging from 113 to 170 GHz. The channel's behavior is scrutinized under conditions of door movements – a scenario commonly triggered by human activity and/or wind. A sophisticated theoretical framework has been established, amalgamating various mechanisms such as transmission, reflection, absorption, and diffraction. This framework leverages the Fresnel formula, a multi-layer transmission model, and the knife-edge diffraction theory. A harmonious alignment between experimental results and theoretical predictions is observed, underscoring the validity of our approach. The findings reveal a correlation between transmitted power loss and both frequency and incident angle, highlighting an increase in power loss at higher frequencies and larger angles. However, it is noteworthy that reflectivity remains invariant with respect to the operating frequency.

When examining the impact of door opening, it is discerned that the edge of the door induces a power blockage of 10 dB, surpassing the 8.3 dB transmission loss attributed to the door glass, yet remaining inferior to the 15 dB blockage caused by metallic door handles. Furthermore, the study elucidates that power fluctuations resulting from door oscillations are inconsequential, even when the amplitude of shaking reaches up to 10 cm.

Finally, this paper provides a comprehensive analysis of terahertz channel performance in indoor settings, highlighting the pivotal role of glass doors in influencing channel behavior. The integration of empirical data and theoretical models offers valuable insights, paving the way for enhanced communication systems and innovative solutions in the terahertz domain.

**Acknowledgements**

This work was supported in part by the National Natural Science Foundation of China (62071046), the Science and Technology Innovation Program of Beijing Institute of Technology (2022CX01023), the Graduate Innovative Practice Project of Tangshan Research Institute, BIT (TSDZXX202201) and the Talent Support Program of Beijing Institute of Technology "Special Young Scholars" (3050011182153).